\documentstyle[psfig,12pt]{article}

\def\beq{\begin{equation}}
\def\eeq{\end{equation}}

\def\be{\begin{equation}}
\def\bea{\begin{eqnarray}}
\def\ee{\end{equation}}
\def\eea{\end{eqnarray}}

\def\d{\partial}
\def\eqref#1{(\ref{#1})}
\def\a{\alpha}
\def\ep{\epsilon}
\def\bra{\langle}
\def\ket{\rangle}
\def\e{{\rm e}}
\def\tr{{\rm tr}}

\def\N#1{{\cal N}=(#1,#1)}
\def\kp{{k^\prime}}

\def\text{\hbox}

\begin{document}


\begin{titlepage}

\begin{centering}

\vspace*{3cm}

{\Large\bf Improved  results for ${\cal N}=(2,2)$ super Yang--Mills theory
using supersymmetric discrete light-cone quantization}

\vspace{1.5cm}

{\bf  Motomichi Harada$^a$, John R. Hiller$^b$, Stephen Pinsky$^a$,
and Nathan Salwen$^a$}

\vspace{0.5cm}

{\sl ${}^a$Department of Physics \\
Ohio State University\\
Columbus OH 43210}

\vspace{0.5cm}

{\sl ${}^b$Department of Physics \\
University of Minnesota Duluth\\
Duluth MN 55812}

\vspace{1.5cm}


\vspace{2cm}

\begin{abstract}
We consider the (1+1)-dimensional ${\cal N}=(2,2)$ super Yang--Mills theory
which is obtained by dimensionally reducing ${\cal N}=1$ super Yang--Mills
theory in four dimension to two dimensions. We do our calculations in the
large-$N_c$ approximation using Supersymmetric Discrete Light Cone
Quantization. The objective is to calculate quantities that
might be investigated by researchers using other numerical methods. We
present a precision study of the low-mass spectrum and the stress-energy
correlator $\bra T^{++}(r) T^{++}(0) \ket$. We find that the mass gap of
this theory closes as the numerical resolution goes to infinity and that the
correlator in the intermediate $r$ region behaves like $r^{-4.75}$.
\end{abstract}

\end{centering}


\vfill

\end{titlepage}

\newpage



\section{Introduction}

There is a pressing need to solve quantum field theories in the
nonperturbative
regime.  Over the last thirty years a significant amount of progress has
been 
made in this area using lattice gauge theory. Many of the most interesting
quantities in QCD and electroweak physics are being calculated to ever
increasing accuracy. There remains, however, a number of nonperturbative
quantities in supersymmetric quantum field theories that are interesting
in a variety of formal physics contexts but have not been calculated. Some
of these calculations are now just beginning to be considered.  With
increased 
interest in the physics of extra dimensions, it is more important
than ever to solve supersymmetric theories in the nonperturbative regime.

The progress in putting supersymmetry on a lattice has been rather slow
due to some critical problems: the lack of translational invariance on a
lattice, the notorious doubling of fermion states~\cite{Nielsen:1980rz}, and
the breakdown of the
Leibniz rule~\cite{Fujikawa:2002ic}. Recently, however, some interesting
new approaches have shed some light on this
issue~\cite{Cohen:2003xe,Cohen:2003qw,Sugino:2003yb,Sugino:2004qd}. These
approaches make possible the restoration of supersymmetry in a continuum
limit 
without fine-tuning of parameters and even without introducing some
``sophisticated'' fermions such as domain-wall~\cite{Kaplan:1992bt} or
overlap fermions~\cite{Narayanan:1994gw,Neuberger:1997fp}. However,
these techniques seem to be applicable to only some subset of all
supersymmetric theories.

Given this increasing interest and promising new ideas for the realization
of supersymmetry on a lattice, it is worthwhile to provide some specific,
detailed numerical results using Supersymmetric Discretized Light Cone
Quantization (SDLCQ)~\cite{sakai95,Lunin:1999ib} for the simplest theory
for which the new lattice techniques are applicable. SDLCQ is a well
established tool for calculations of physical quantities in supersymmetric
gauge theory and has been exploited for many supersymmetric Yang--Mills
(SYM) 
theories. The ${\cal N}=(2,2)$ theory in 1+1 dimensions in the large-$N_c$
limit is discussed in Ref.~\cite{Antonuccio:1998mq}; however, the published
results are primitive compared to what can be obtained today because of our
greatly improved hardware and software.  In this paper we are now able to
reach a resolution of $K=12$, while in Ref.~\cite{Antonuccio:1998mq} we
could 
reach only $K=5$. Here we will present new and more detailed results on
this theory against which the lattice community can compare the results of
their new techniques.

Briefly, the SDLCQ method rests on the ability to produce an exact
representation of the superalgebra but is otherwise very similar to Discrete
Light Cone Quantization (DLCQ)~\cite{pb85,bpp98}. In DLCQ we compactify the
$x^-$ direction by putting the system on a circle with a period
of $2L$, which discretizes the longitudinal momentum as $p^+=n\pi/L$, where
$n$ is an integer. The total longitudinal momentum $P^+$ becomes $K\pi/L$,
where $K$ is an integer called the harmonic resolution~\cite{pb85}. The
positivity of the light-cone longitudinal momenta then limits the number of
possible Fock states for a given $K$, and, thus, the dimension of Fock space
becomes finite, enabling us to do some numerical computations. It is assumed
that as $K$ approaches infinity, the solutions to this large finite problem
approach the solutions of the field theory. The difference between DLCQ and
SDLCQ lies in the choice of discretizing either $P^-$ or $Q^-$ to construct
the matrix approximation to the eigenvalue problem
$M^2|\Psi\ket=2P^+P^- |\Psi\ket=2P^+(Q^-)^2/\sqrt 2|\Psi\ket$, with 
$P^+=K\pi/L$. For more details and additional discussion of SDLCQ, we refer 
the reader to Ref.~\cite{Lunin:1999ib}.

An interesting new result of the calculation we present here is that
finite-dimensional representations of the SDLCQ  with
odd and even values of $K$ result in very distinct solutions
of the $\N 2$ SYM theory, which only become identical as $K$ approaches
infinity.  One might initially think that this is a shortcoming of the SDLCQ
approach, but it turns out to be an advantage because it provides an
internal 
measure of convergence.

We will give some numerical results of the low-energy 
spectrum. There we will see that as we go to higher and higher resolutions, 
we find  bound states with lower and lower mass. We have seen this behavior in the
${\cal N}=(1,1)$ theory where the lowest mass state converges linearly to
zero as a function of $1/K$.  This closing of the mass gap as
$K\to \infty$ was predicted by Witten~\cite{Witten:1995im} for the
${\cal N}=(1,1)$ and ${\cal N}=(2,2)$ theories.
We find that 
in the latter case the convergence is not linear in $\frac{1}{K}$,
and, while our results are
consistent with the mass gap going to zero, they are not conclusive.

We have also been able to solve analytically for the wave functions of some
of the pure bosonic massless states, and we will present the exact form of the wave
function for some cases. We will show that the states must have certain 
properties to be massless, which then enable us to count the number of the 
states for a given resolution $K$. In addition, we will present the 
formulae to count a minimum total number of massless states. 

Finally, we will look at the two-point correlation function of the
stress-energy tensor $\bra T^{++}(r)T^{++}(0)\ket $.
We see the expected $1/r^4$-behavior in the UV and IR regions,
and, interestingly, we find that the correlator behaves as $1/r^{4.75}$ in
the intermediate region. We know of no predictions for this behavior;
however, 
for $\N 8$ SYM theory there is a prediction that this correlator should
behave like $1/r^5$ in the intermediate region.
 
The structure of this paper is the following. In Sec.~\ref{sec:N2SYM} we
focus our attention on the low-energy states. After giving a quick
review of $\N 2$ SYM theory with SDLCQ, we give some
numerical results for the low-energy states, discuss analytically some
properties of pure bosonic massless states, and present the formulae to count 
a minimum total number of massless states. We discuss the numerical
results for the two-point correlation function of the stress-energy tensor
in Sec.~\ref{sec:cor}. A summary and some additional discussion are given in
Sec.~\ref{sec:discussion}.


\section{Review of ${\cal N}$=(2,2) SYM theory} \label{sec:N2SYM}


\subsection{${\cal N}$=(2,2) SYM theory and SDLCQ}

Before giving the numerical results, let us quickly review some analytical
work on ${\cal N}$=(2,2) SYM theory for the sake of completeness. For more
details see Ref.~\cite{Antonuccio:1998mq}.
This theory is obtained by dimensionally reducing ${\cal N}$=1 SYM
theory from four dimensions to two dimensions. In light cone gauge, where
$A_-=0$, we find for the action
\begin{eqnarray}
S^{LC}_{1+1}&=&\int dx^+ dx^- \tr \Bigg[ \d_+ X_I \d_-X_I
 +i\theta^T_R \d^+\theta_R+i\theta^T_L\d^-\theta_L  \\
 &&\quad +\frac 12 (\d_-A_+)^2+gA_+J^++\sqrt 2 g\theta^T_L\ep_2\beta_I
 [X_I,\theta_R]+\frac {g^2}4 [X_I,X_J]^2 \Bigg], \nonumber
\end{eqnarray}
where $x^{\pm}$ are the light-cone coordinates in two dimensions,
the trace is taken over the color indices, $X_I$ with $I=1,2$ are
the scalar fields and the remnants of the transverse components of the
four-dimensional gauge field $A_{\mu}$, two-component spinor fields
$\theta_R$ 
and $\theta_L$ are remnants of the right-moving and left-moving projections
of the four-component spinor in the four-dimensional theory, and $g$ is the
coupling constant. We also define the current
$J^+=i[X_I,\d_-X_I]+2\theta^T_R\theta_R$,
and use the Pauli matrices
$\beta_1\equiv\sigma_1$, $\beta_2\equiv\sigma_3$, and $\ep_2\equiv
-i\sigma_2$.

After eliminating all the non-dynamical fields using the equations of
motion, 
we find for $P^{\a}=\int dx^- T^{+\a}$
\be 
P^+=\int dx^- \tr(\d_-X_I\d_-X_I+i\theta_R^T\d_-\theta_R),
\ee
and
\be
P^-=g^2\int dx^- \tr\left(-\frac 12 J^+\frac 1{\d^2_-}J^+
   -\frac 14[X_I,X_J]^2+\frac i2 (\ep_2\beta_I[X_I,\theta_R])^T\frac 1{\d_-}
   \ep_2\beta_J[X_J,\theta_R]\right).
\ee
The supercharges are found by dimensionally reducing the supercurrent in
the four-dimensional theory. They are
\be
Q^+_{\a}=2^{5/4}\int dx^- \tr(\d_-X_I\beta_{I\a \eta}u_{\eta}),
\ee
\be
Q^-_{\a}=g\int dx^- \tr\left( -2^{3/4}J^+\frac
1{\d_-}\ep_{2\a\eta}u_{\eta}+2^{-1/4}i[X_I,X_J](\beta_I\beta_J\ep_2)_{\a\eta
}u_{\eta}\right),
\ee 
where $\a,\eta=1,2$ and $u_{\a}$ are the components of $\theta_R$.

We expand the dynamical fields $X_I$ and $u_{\a}$ in Fourier modes as
\be X_{Ipq}(x^-)=\frac 1{\sqrt{2\pi}}\int_0^{\infty}
\frac {dk^+}{\sqrt{2k^+}}[A_{Ipq}(k^+)\e^{-ik^+x^-}
 +A^{\dag}_{Iqp}(k^+)\e^{ik^+x^-}],
\ee
\be 
u_{\a pq}(x^-)=\frac 1{\sqrt{2\pi}}\int_0^{\infty}
   \frac {dk^+}{\sqrt{2}}[B_{\a pq}(k^+)\e^{-ik^+x^-}
   +B^{\dag}_{\a qp}(k^+)\e^{ik^+x^-}],
\ee
where $p,q=1,2,\ldots, N_c$ stand for the color indices, and
$A,B$ satisfy the usual commutation relations
\begin{eqnarray} 
[A_{Ipq}(k^+),A^{\dag}_{Jrs}(k^{'+})]
   =\delta_{IJ}\delta_{pr}\delta_{qs}\delta(k^+-k^{'+}), \\
   \{B_{\a pq}(k^+),B^{\dag}_{\beta rs}(k^{'+})\}
   =\delta_{\a\beta}\delta_{pr}\delta_{qs}\delta(k^+-k^{'+}).
\end{eqnarray}
We work in a compactified $x^-$ direction of length $2L$ and ignore
zero modes.  With periodic boundary conditions we restrict to
a discrete set of momenta
~\cite{sakai95} 
\begin{equation}
  k^+ = \frac{\pi}{L} k, \quad k = 1,2,3,\ldots,\quad
  \int dk^+ \rightarrow \frac{\pi}{L}\sum_{k=1}^{\infty},
  \quad \delta(k^+ - k^{\prime +}) \rightarrow\frac{L}{\pi} \delta_{k\kp}
\end{equation}
Relabeling the operator modes
$\sqrt{\frac{L}{\pi}}a(k) = A(k^+ = \frac{\pi k}{L})$ and
$\sqrt{\frac{L}{\pi}}b(k) =  B(k^+ = \frac{\pi k}{L})$, so that
\begin{eqnarray}
  [a_{Ipq}(k),a^{\dag}_{Jrs}(\kp)]
   =\delta_{IJ}\delta_{pr}\delta_{qs}\delta_{k\kp}, \quad
   \{b_{\a pq}(k),b^{\dag}_{\beta rs}(\kp)\}
   =\delta_{\a\beta}\delta_{pr}\delta_{qs}\delta_{k\kp}.
\end{eqnarray}
the expansion is
\be 
\label{eq:discretized}
X_{Ipq}(x^-)=\frac 1{\sqrt{2\pi}}\sum_{k=1}^{\infty}
\frac {1}{\sqrt{2k}}[a_{Ipq}(k)\e^{-i\frac{\pi}{L}kx^-}
 +a^{\dag}_{Iqp}(k^+)\e^{i\frac{\pi}{L}kx^-}],
\ee
\be
\label{eq:discretized2} 
u_{\a pq}(x^-)=\frac 1{\sqrt{2L}}\sum_{k=1}^{\infty}
   \frac {1}{\sqrt{2}}[b_{\a pq}(k)\e^{-i\frac{\pi}{L}kx}
   +b^{\dag}_{\a qp}(k)\e^{i\frac{\pi}{L}kx^-}].
\ee

In terms of $a$ and $b$, the supercharges are given by
\be
Q_{\a}^+=2^{1/4}i\sqrt{\frac{\pi}{L}}\sum_{k=1}^{\infty}
\sqrt k \beta_{I\a\eta}[a^{\dag}_{Iij}
   (k)b_{\eta ij}(k)-b^{\dag}_{\eta ij}(k)a_{Iij}(k)],
\ee
and
\begin{eqnarray}
   Q_{\a}^-
   &=&\frac{i2^{-1/4}g}{\pi}\sqrt{\frac{L}{\pi}}\sum^{\infty}_{k_1,k_2,k_3
   = 1}
    \delta_{(k_1+k_2),k_3} \Biggl\{ (\epsilon_2)_{\a\eta} \\
   &&\times \Biggl[ \frac 1{2\sqrt{k_1k_2}}\left(\frac {k_2-k_1}{k_3}\right)
    [b^{\dag}_{\eta ij}(k_3)a_{Iim}(k_1)a_{Imj}(k_2)-a^{\dag}_{Iim}(k_1)
    a^{\dag}_{Imj}(k_2)b_{\eta ij}(k_3)] \nonumber \\
   &&+ \frac 1{2\sqrt{k_1k_3}}\left(\frac {k_1+k_3}{k_2}\right)
    [a^{\dag}_{Iim}(k_1)b^{\dag}_{\eta mj}(k_2)a_{I ij}(k_3)-a^{\dag}_{I ij}
    (k_3)a_{Iim}(k_1)b_{\eta mj}(k_2)] \nonumber \\
   &&+ \frac 1{2\sqrt{k_2k_3}}\left(\frac {k_2+k_3}{k_1}\right)
    [a^{\dag}_{I ij}(k_3)b_{\eta im}(k_1)a_{Imj}(k_2)-b^{\dag}_{\eta
im}(k_1)
    a^{\dag}_{Imj}(k_2)a_{I ij}(k_3)] \nonumber \\
   &&-\frac 1{k_1}[b^{\dag}_{\eta ij}(k_3)b_{\eta im}(k_1)b_{\eta mj}(k_2)
    +b^{\dag}_{\eta im}(k_1)b^{\dag}_{\eta mj}(k_2)b_{\eta ij}(k_3)]
    \nonumber \\
   &&-\frac 1{k_2}[b^{\dag}_{\eta ij}(k_3)b_{\eta im}(k_1)b_{\eta mj}(k_2)
    +b^{\dag}_{\eta im}(k_1)b^{\dag}_{\eta mj}(k_2)b_{\eta ij}(k_3)]
    \nonumber \\
   &&+\frac 1{k_3}[b^{\dag}_{\eta ij}(k_3)b_{\eta im}(k_1)b_{\eta mj}(k_2)
    +b^{\dag}_{\eta im}(k_1)b^{\dag}_{\eta mj}(k_2)b_{\eta ij}(k_3)]
    \Biggr] \nonumber \\
    && + 2  (\epsilon_2)_{IJ}
    \Biggl( \frac 1{4\sqrt{k_1k_2}}
    [b^{\dag}_{\alpha ij}(k_3)a_{I im}(k_1)a_{J mj}(k_2)
    +a^{\dag}_{J im}(k_1)a^{\dag}_{I mj}(k_2)b_{\alpha ij}(k_3)] \nonumber
\\
   &&+\frac 1{4\sqrt{k_2k_3}}
    [a^{\dag}_{J ij}(k_3)b_{\alpha im}(k_1)a_{I mj}(k_2)
    +b^{\dag}_{\alpha im}(k_1)a^{\dag}_{J mj}(k_2)a_{I ij}(k_3)] \nonumber
\\
   &&+\frac 1{4\sqrt{k_3k_1}}
    [a^{\dag}_{I ij}(k_3)a_{J im}(k_1)b_{\alpha mj}(k_2)
    +a^{\dag}_{I im}(k_1)b^{\dag}_{\alpha mj}(k_2)a_{J ij}(k_3)]
    \Biggr)\Biggr\}.\nonumber
\end{eqnarray}    
using the relation 
$([\beta_I,\beta_J]\epsilon_2)_{\alpha\eta} =
\delta_{\alpha\eta} (\epsilon_2)_{IJ}$.

They satisfy the superalgebra conditions for anticommutators involving
$Q^+_{\a}$,
\begin{equation}
   \{Q^{+}_{\a},Q^{+}_{\beta}\}=\delta_{\a\beta}2\sqrt 2 P^{+}, \quad
   \{Q^+_{\a},Q^-_{\beta}\}=0. \label{superalgebra} 
\end{equation}
but do not satisfy the the condition
$\{Q^{-}_{\a},Q^{-}_{\beta}\}=\delta_{\a\beta}2\sqrt 2 P^{-}$.
Instead, in SDLCQ we find
\be
\{Q^{-}_{\a},Q^{-}_{\beta}\}\ne 0 \ {\rm if} \ \a\ne\beta, \qquad
    (Q^-_1)^2=\sqrt 2 P^-_1 \ne \sqrt 2 P^-_2=(Q^-_2)^2.
\ee
Although we have different $P^-_{\a}$ for different $Q^-_{\a}$, we can
define a unitary, self-adjoint transformation $C$, such that
\be 
C a_{1 ij} C  = a_{2 ij}, \quad
   C b_{1 ij} C = -b_{2 ij}.
\ee
and find that $C P^-_1 C = P^-_2$.  Thus the eigenvalues of
$P^-_{\a}$ 
are the same. 
We may choose either one of the two
$Q^-_{\a}$'s, at least for our purposes, and in what follows we will use
$Q^-_1$ and will suppress the subscript unless it is needed for clarity.

The momentum, $P^+$, is given by
\begin{equation}
  P^+ = \frac{1}{\sqrt{2}} (Q^+_1)^2 =
  \frac{\pi}{L}
  \sum_{k}
  k
  \bigl(
  a_{Iij}^{\dag} a_{Iij} +
  b_{\nu ij}^{\dag} b_{\nu ij} \bigr)
\end{equation}
We work with a fixed value of momentum
\begin{eqnarray}
  P^+ = \frac{\pi}{L} K, \quad K = 1,2,\ldots
\end{eqnarray}
We call $K$ the resolution because larger values of $K$ allow larger
values of $L$ while leaving the momentum $P^+$ fixed.

The next thing to note is that there are three $Z_2$ symmetries of
$Q_1^-$. 
The first one is $R_1$-symmetry, where $R_{\alpha}$ acts as follows
\be 
a_{1ij} \leftrightarrow a_{2ij}, \quad
   b_{\a}\rightarrow -b_{\a}
\ee
The second is $S$-symmetry
\be 
a_{I ij} \rightarrow -a_{I ji}, \quad
   b_{\a ij}\rightarrow -b_{\a ji}.
\ee
The third is what we call $T$-symmetry
\be
a_{Iij}\rightarrow -a_{Iij}, \quad b_{\a} \ {\rm unchanged}.
\ee
It is easy to see that under these symmetries $Q^-_1$ is invariant.

Using the relations,
\[ R_1 Q^+_{1} R_1 = -Q_{2}^+, \quad TQ^+_{\a}T = - Q^+_{\a},\]
we find
\begin{eqnarray}
  R_1 (Q^+_{1} \pm Q_{2}^+) R_1 = \mp (Q^+_{1} \pm Q_{2}^+), \quad
  T (Q^+_{1} \pm Q_{2}^+) T = - (Q^+_{1} \pm Q_{2}^+).
\end{eqnarray}
Also note that
\begin{eqnarray}
  \{Q^+_{1} \pm Q_{2}^+,Q^+_{1} \pm Q_{2}^+\}
  =
  \{Q^+_{1} ,Q^+_{1} \} +
  \{ Q_{2}^+, Q_{2}^+\} +
  \pm 2   \{Q^+_{1} , Q_{2}^+\} = 4 \sqrt{2} P^+.
\end{eqnarray}
We work in a subspace of definite momentum so $(Q^+_{1} \pm
Q_{2}^+)$ must have non zero eigenvectors.
Since $Q^+_{\a}$ and $Q^-_{\a}$ are fermionic operators we see that
a bosonic energy eigenstate
$|\Psi_B\ket_{++}$ which is even under $R$ and $T$-symmetry,
can be transformed into
\be 
  |\Psi_B\ket_{\mp -}=Q^-_1(Q^+_1\pm Q_2^+)|\Psi_B \ket_{++}, \quad
  |\Psi_B\ket_{-+}=(Q^+_1+Q^+_2)(Q^+_1-Q^+_2)|\Psi\ket_{++}
\ee
which are all degenerate with $|\Psi_B\ket_{++}$.
One should notice here that
we cannot use $Q^-_1$ and $Q_2^-$ at the same time since they do not commute
with each other. Thus, including the supersymmetry, we have an 8-fold
degeneracy. Utilizing the remaining $S$-symmetry, which does not
give us a mass degeneracy, we can divide the mass spectrum into 16
independent 
sectors. 
This significantly reduces the size of the computational problem. It
will be convenient to refer to bound states of this theory as having $S$,
$T$, 
or $R$ even or odd parity and to refer to a state as
having even or odd resolutions if $K$ is an even or odd integer.


\subsection{Mass gap}

Tables~\ref{mass+} and \ref{mass-} show the first few low-mass states. We
find anomalously light states in the sectors with opposite
$K$ and $S$ parity  for $K$ larger than 4. Furthermore, the number of
extremely light states increases by one as we increase $K$ by two. We
believe 
that these anomalously light states should be exactly massless states, but
for 
some reason there is an impediment preventing SDLCQ from achieving this
result.
Some of the evidence for this comes from a study of the
average number of partons $\bra n \ket$ in the bound states. For example, in
the sector with $S$ and $K$ even, for each even integer $r$ less than $K$,
there is exactly one bosonic massless state with $\bra n \ket=r$. For $K$
odd 
we do not see massless states of this type, but we do find
$\bra n \ket=r$ for the anomalously light bound states in this sector. This
is also the first sign of the distinction between representations of the
supersymmetry algebra in different symmetry sectors, namely those
with anomalously 
light states (with opposite $S$ and $K$ parity) and those without
anomalously 
light states (with matching $S$ and $K$ parity).
\begin{table}[t]
\begin{tabular}{|cccccccccc|}
\hline 
$K$=3&4 & 5& 6& 7& 8& 9& 10& 11& 12 \\
\hline
1.308&4.009&0.0067&2.144&0.0040&1.415&0.0026&1.040&0.0018&0.8188 \\

12.62&12.24&0.6304&2.514&0.0060&1.5999&0.0038&1.138&0.0026&0.8790 \\

22.06&15.04&1.0813&2.645&0.4366&1.712&0.0048&1.212&0.0026&0.9312 \\

      &15.28&1.1099&2.773&0.6016&1.729&0.3515&1.256&0.0039&0.9397 \\

      &22.53&1.5732&2.807&0.6308&1.811&0.4372&1.347&0.3062&1.0072 \\
\hline
\end{tabular}
\caption{The mass squared $M^2$ of the first few lowest massive states in the $S$-even sector in units of $g^2N_c/\pi$ for a series of resolutions $K$.}
\label{mass+}
\end{table}
\begin{table}[t]
\begin{tabular}{|ccccccccc|}
\hline
 $K$=4 & 5& 6& 7& 8& 9& 10& 11& 12 \\
\hline
1.2009&3.1876&0.00674&1.8427&0.00440&1.2687&0.00302&0.95786&0.00217 \\

1.2009&3.1887&0.6402 &1.9305&0.00538&1.3266&0.00317&0.99795&0.00218 \\

12.296&3.3239&0.6747 &2.0413&0.45529&1.4087&0.00431&1.0302 &0.00219 \\

12.296&11.489&0.9900 &2.1415&0.48010&1.5107&0.36858&1.1036 &0.00356 \\

19.502&11.492&1.0313 &2.3603&0.55873&1.5219&0.38647&1.1345 &0.32053 \\
\hline
\end{tabular}
\caption{Same as Table 1 but for the $S$-odd sector.}
\label{mass-}
\end{table}

\begin{figure}[ht]
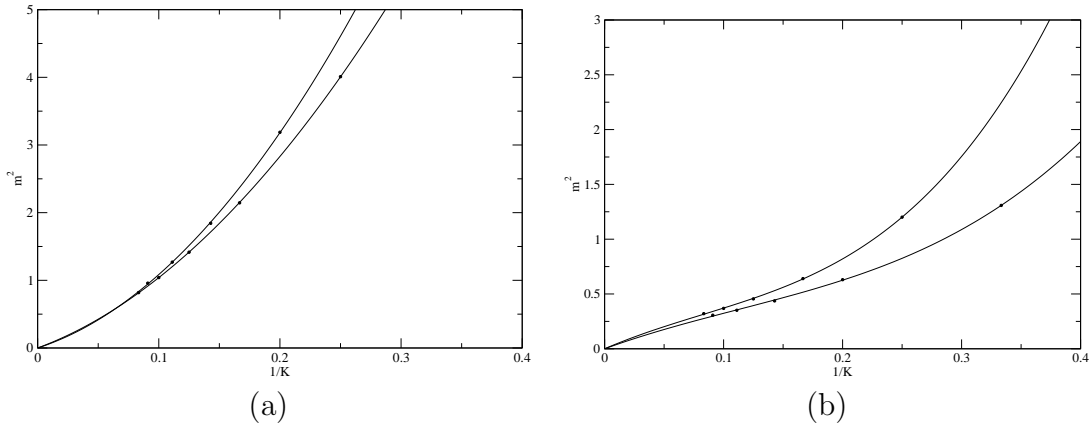

\begin{center}
\begin{tabular}{cc}
\psfig{figure=statehigh.eps,width=7cm} &
\psfig{figure=statelow.eps,width=7cm}
\\
(a)&(b)
\end{tabular}
\end{center}
\caption{Plots of the mass squared in units of $g^2N_c/\pi$ for the lowest massive states, excluding the anomalously light states, with a polynomial fit constrained to go through the origin. The plot in (a) corresponds to the sector where $S$ and $K$ have the same parity, and the plot in (b) to the sector where $S$ and $K$ have opposite parity.}
\label{low}
\end{figure}

In our discussion of the mass gap we will not include the anomalously light
states as part of the massive spectrum for the reason given above. To study
the mass gap we will look at the lowest massive state in each sector as a
function of $1/K$ as shown in Fig.~\ref{low}.
There we also show polynomial fits in all four sectors separately.
The fits are constrained to go through the origin. The quadratic fits
look very good in Fig.~\ref{low}(a). but Fig.~\ref{low}(b) required
a cubic.  The two fits with opposite $S$ and $K$ parity look very similar
as do the two fits with same $S$ and $K$ parity. In each case we could have
fit all the points with one curve if we were to include a small oscillatory
function in the fit. We should note here that oscillatory behavior has
been observed before in different
theories~\cite{Gross:1997mx,Hiller:2003qe}.
The explanation given there is that those states which show
the oscillatory behavior comprise non-interacting two-body states. This,
however, does not seem applicable in our case
since the states in Fig.~\ref{low}
are the lowest energy states; thus there are no lower energy states
available
to form two-body states.

The distinct character of the mass gap
serves as another piece of evidence that we have two different
classes of representations.
The data is consistent with the mass gap closing to $0$ as $K\to
\infty$, especially for the case where $S$ and $K$ have the same parity.
The odd and even representations approach each other as $K$ increases
and we hypothesize that they become identical
in the continuum limit of
$K\to \infty$. When we present the correlation function in
Sec.~\ref{sec:cor}, we will see further evidence for this claim.


\subsection{Massless states}


\subsubsection{Pure bosonic massless states}

Let us investigate the properties of pure bosonic massless states in full 
detail in the $N_c \to \infty$ limit. This is done by generalizing the 
discussion of the bound states in SDLCQ for ${\cal N}$=(1,1) SYM theory,
as given in Refs.~\cite{Lunin:1999ib,Antonuccio:1998kz}, to 
${\cal N}$=(2,2) SYM theory. 

For simplicity, let us consider the states consisting of a fixed $n$ number 
of partons only. A pure bosonic massless state is given by
\[
|\Psi,0\ket=N\sum_{q_1,\ldots,q_n}\sum_A
   \delta_{(q_1+\ldots+q_n),K} \bar f^{(0)}_{[A_1\ldots A_n]}(q_1\ldots q_n)
   \tr[a_{A_1}^{\dag}(q_1)\ldots a_{A_n}^{\dag}(q_n)] |0\ket, 
\]
where $N$ is the normalization factor, $q_i=1,2,\ldots$ is the unit of the 
light-cone momentum $p_i=q_i\pi/L$ carried by the $i$-th parton, $A_i=1,2$ 
indicates the flavor index for each parton, the sum 
$\sum_A$ is the summation over all possible permutations of the flavor indices 
$A_i$'s, $\bar f$ is the wave function, and the trace is taken over the color indices. 
Note that we don't have the symmetry factor coming from 
the cyclic property of the trace in the above notation; one has to put in the 
symmetry factor by hand if one would like it to be in there as we will do so 
for an example given later in this subsection. In other words, Fock states with non-zero symmetry factor are not normalized. 

Due to the cyclic property of the trace, we have 
\[ \bar f_{[A_1\ldots A_{n}]}(q_1,\ldots,q_n)
   =\bar f_{[A_2\ldots A_{n}A_1]}(q_2,\ldots,q_n,q_1)=\ldots =
   \bar f_{[A_nA_1\ldots A_{n-1}]}(q_n,\ldots,q_{n-1}). 
\]

Since $P^-=(Q^-)^2/\sqrt 2$, all the massless states should vanish upon the 
action of $Q^-$.  Thus, we must have $Q^-|\Psi,0\ket=0$. This identity, 
however, can be simplified somewhat for pure bosonic massless states. That is, 
the terms to consider in $Q^-$ are those which annihilate one boson and create 
one boson and one fermion, and those which annihilate two bosons and create 
one fermion. Both the former and latter class of terms in $Q^-$ separately 
annihilates $|\Psi,0\ket$. In the large-$N_c$ limit the former class gives, 
writing $f(q_1,\ldots,q_n)\equiv \sqrt{q_1\ldots q_n}\bar f(q_1,\ldots,q_n)$,
\begin{eqnarray}
&0&=(\ep_2)_{\a\beta}\Bigl\{
\frac{2q_{n-1}+t}{(q_{n-1}+t)t}f^{(0)}_{[A_1\ldots A_n]}
(q_1,\ldots,q_{n-1}+t,q_n) \nonumber \\
&&\quad 
-\frac{2q_{n}+t}{(q_{n}+t)t}f^{(0)}_{[A_1\ldots A_n]}
(q_1,\ldots,q_{n-1},q_n+t)\Bigl\} \nonumber \\
&&
\quad +\frac {M_{IA_n}^{\a\beta}}{2(q_n+t)}
f^{(0)}_{[A_1\ldots A_{n-1},I]}(q_1,\ldots,,q_{n-1},q_n+t) \nonumber \\
&&
\quad -\frac {M_{IA_{n-1}}^{\a\beta}}{2(q_{n-1}+t)}
f^{(0)}_{[A_1\ldots A_{n-2},I,A_n]}(q_1,\ldots,,q_{n-1}+t,q_n), \label{eq1}
\end{eqnarray}
and the latter yields
\begin{equation}
0=\sum_{A_{n-1},A_n}\sum_k \left( (\ep_2)_{\a\beta} \frac{t-2k}{tk(t-k)}
     \delta_{A_{n-1},A_n}+\frac {M_{A_{n-1}A_n}^{\a\beta}}{k(t-k)}\right)
     f^{(0)}_{[A_1\ldots A_n]}(q_1,\ldots,q_{n-2},k,t-k), \label{eq2}
\end{equation}
where 
$M_{IJ}^{\a\beta}\equiv [(\beta_I\beta_J-\beta_J\beta_I)\epsilon_2]_{\a\beta}$, 
$t$ is the momentum of the created fermion, and the momentum conserving 
Kronecker's delta $\delta_{(q_1+\ldots+q_n),K}$ is understood implicitly. These 
are the necessary and sufficient conditions for a pure bosonic state to be massless. 
One should notice that the above equations reduce to the corresponding 
equations found in Ref.~\cite{Lunin:1999ib,Antonuccio:1998kz} with 
$(\ep_2)_{\a\beta}=1$, $A_i=1$ for all $i$'s, and $M_{IJ}^{\a\beta}=0$, as 
expected. 

In principle, we could find the properties of all kinds of pure bosonic 
massless states using Eqs.~(\ref{eq1}) and (\ref{eq2}). However, we limit 
ourselves here to the investigation of only two special types. To simplify
the notation, we omit the superscript $(0)$ from the wave function $f$ hereafter.

The simplest case is where $n=K$, that is to say, all the partons have one 
unit of momentum $\pi/L$ and, thus, $f=\bar f$. In this case Eq.~\eqref{eq1} is 
trivially satisfied since we cannot have states with $(K+1)$ partons. From 
Eq.~\eqref{eq2} we get 
\begin{equation}
 0=f_{[A_1\ldots A_{n-2},1,2]}-f_{[A_1\ldots A_{n-2},2,1]} ,\label{1eq1}
\end{equation}
where we have omitted $(q_1,\ldots,q_n)=(1,\ldots,1)$. Eq.~\eqref{1eq1} means, 
with the help of the cyclic property of $f$, that the wave function is 
unchanged after moving {\em any} flavor index 
to {\em any} location in the list of indices. For instance, we find, writing 
$f_{[A_1\ldots A_{n}]} \equiv [A_1\ldots A_n]$, 
\[ [1212]=[1221]=[2121]=[2211]=[2112]=[1122] . \]
It is clear that the state with the above six wave functions being the same 
and all others zero satisfies \eqref{1eq1}, or equivalently 
Eqs.~(\ref{eq1}) and (\ref{eq2}), the necessary and
sufficient conditions to be massless. Therefore, writing 
$\tr[a^{\dag}_{A_1}(1)\ldots a^{\dag}_{A_n}(1)]|0\ket \equiv A_1\ldots A_n$, 
we find the state 
\[ N[1212](1212+1221+2121+2211+2112+1122)=N[1212](2(1212)+4(1122)) \]
is massless, where we used the cyclic property of $f$. In terms of the 
normalized Fock states $\tr[a^{\dag}_{A_1}(1)\ldots a^{\dag}_{A_n}(1)]
|0\ket/(\sqrt s N_c^{n/2}) \equiv \underline{A_1\ldots A_n}=A_1\ldots 
A_n/(\sqrt s N_c^{n/2})$, 
where $s$ is the symmetry factor, we find, after normalizing properly, that
\[ \frac 1{\sqrt 3}(\underline{1212})+\sqrt {\frac 23}(\underline{1122})  \]
is massless since $s$ for 1212 and 1122 equals two and one, respectively. 
Indeed we have found the very same massless state in our 
numerical results. 

As we have seen above, there is a one-to-one correspondence between a 
massless state and a given set of flavor indices, which has a {\em fixed} 
number of 1's and 2's. This means that every time we change the number of 1's 
(or 2's) in the flavor indices, we find a new massless state.  Since we can 
have $K+1$ such different sets of flavor indices, we have $K+1$ massless states
of this kind. As verification of our argument, we enumerated all the massless 
states for $K$ up to six and found all of them with the correct coefficients 
in our numerical results. 

The next case to consider is where $n=K-1$. In this case only one of the partons 
has two units of momentum, so that $f=\sqrt 2 \bar f$. However, since all the $f$'s 
have the same factor of $\sqrt 2$, we can absorb $\sqrt 2$ into the normalization 
factor $N$ and practically can set $f\equiv \bar f$. We have $t=1$ and 
$q_i=1$ with $i=1,\ldots,n$ in Eq.~\eqref{eq1} and find, writing 
$(q_1,\ldots,q_n)=(1,\ldots,1,1,2)\equiv (1,2)$ and so on,  
\begin{equation}
 0=[A_1\ldots A_{n}](2,1)-[A_1\ldots A_{n}](1,2) ,\label{2eq1}
\end{equation}
\begin{equation}
 0=[A_1\ldots A_{n-2},A_{n-1},A_n](1,2)-[A_1\ldots A_{n-2},A_n,A_{n-1}](2,1),
 \label{2eq2}
\end{equation}
\begin{equation}
 0=[A_1\ldots A_{n-2},A_{n-1},A_{n-1}](1,2)+[A_1\ldots A_{n-2},A_n,A_n](2,1),
 \label{2eq3}
\end{equation}
where $A_{n-1}\ne A_n$ in Eqs.~(\ref{2eq2}) and (\ref{2eq3}). 
For Eq.~\eqref{eq2} we have $t=2$, $k=1,2$ and $q_i=1$ with $i=1,\ldots,n-2$, 
and we get
\begin{equation}
 0=[A_1\ldots A_{n-2},A,A](1,2)-[A_1\ldots A_{n-2},A,A](2,1) ,\label{2eq4}
\end{equation}
\begin{eqnarray}
 0&=&[A_1\ldots A_{n-2},1,2](1,2)+[A_1\ldots A_{n-2},1,2](2,1) \nonumber \\
 &&\quad -[A_1\ldots A_{n-2},2,1](1,2)-[A_1\ldots A_{n-2},2,1](2,1).
 \label{2eq5}
\end{eqnarray}
Apparently, we have five equations for the massless states to satisfy, but 
it is easy to see that Eq.~\eqref{2eq4} is incorporated into Eq.~\eqref{2eq1} 
and that if Eqs.~(\ref{2eq1}) and (\ref{2eq2}) are true, so is Eq.~\eqref{2eq5}
automatically. Hence, the three equations Eqs.~(\ref{2eq1}), (\ref{2eq2}),
and (\ref{2eq3}) are in fact the equations for
massless states to satisfy for $n=K-1$. 

In order to see what the three equations allow us to do, let us first write 
\[ [A_1,\ldots,A_n](1,2)\equiv [A_1,\ldots,A_n'] .  \]
That is, let us put a prime on 
top of an index whose corresponding parton has two units of momentum. Then, 
Eq.~\eqref{2eq1} allows us to move the ``prime" to any index. 
Eq.~\eqref{2eq2}, along with this fact, then also allows us to
move the index with a prime to any location in the index list. For example,
we have 
\[  [112']=[11'2]=[1'12]=[12'1]=[1'21]=[121']=[2'11]=[21'1]=[211']. \]
Furthermore, Eq.~\eqref{2eq3} allows us to replace $11'$ by $2'2$ (or $22'$ 
using Eq.~\eqref{2eq1}) as long as a minus sign is inserted. Thus, for the 
above example we get
\[ [112']=[11'2]=[1'12]=-[22'2], \]
where we have omitted the wave functions related by cyclic permutations. This
means that the state 
\[ (112'+11'2+1'12-22'2)/2 \]
is massless. Note that the symmetry factor in this case is equal to one for 
all the Fock states above.

Since Eqs.~(\ref{2eq1}), (\ref{2eq2}), and (\ref{2eq3}) relate all the 
sets of flavor indices with an even/odd number of 1's to one another, we 
have only {\em two} independent sets of flavor indices: the one with even 
numbers of 1's and the other with an odd number. This means that there are 
{\em two} massless states of this type. Again we have confirmed this statement 
numerically for $K$ up to six. 

To summarize, we have found in the large-$N_c$ limit the necessary and 
sufficient conditions, Eqs.~(\ref{eq1}) and (\ref{eq2}), that pure bosonic 
massless states are to satisfy. As an application we considered two special 
cases and found that there are $K+1$ massless states of the
type $\tr[a^{\dag}_{A_1}(1)\ldots a^{\dag}_{A_K}(1)]$ and two of the type
$\tr[a^{\dag}_{A_1}(1)\ldots a^{\dag}_{A_{K-2}}(1)a^{\dag}_{A_{K-1}}(2)]$. 
Also, we gave a way to enumerate all such massless states for a given $K$.

\subsubsection{Count of massless states}

It is possible to predict a minimum number of massless states by
comparing the number of states in the different symmetry sectors.
Since $(Q^-)^2$ takes a state from one symmetry sector to another and
then back it must have $0$ eigenvalues if the dimensionality of the
intermediate sector is less than that of the original sector.  It is
possible to create a simple recursive formula for the number of states
in each sector\cite{ToAppear}.  For the case when $K$ is prime and
odd, the formula is particularly simple. We present the results here
but refer to the other publication for justification. 
We define $A_{bes^+}(K,n)$ as the number of states in the bosonic sector 
with an even number of partons and even $S$ symmetry, where $n$ indicates 
how many types of particles we have in a SYM theory, i.e. $n=4$ for 
${\cal N}=(2,2)$ SYM.  Then
\begin{eqnarray}
  &&A_{bes^+}(K,n) = A_{fes^+}(K,n) = \frac{A_f(K,n) + A_f(K,-n) + W}{2} \\ 
  \nonumber
 && A_{bes^-}(K,n) = A_{fes^-}(K,n) = \frac{A_f(K,n) + A_f(K,-n) - W}{2} \\
  \nonumber
 && A_{bos^+}(K,n) = A_{fos^+}(K,n) = 
  A_{bos^-}(K,n) = A_{fos^-}(K,n) =
  \frac{A_f(K,n) - A_f(K,-n)}{2} 
\end{eqnarray}
where
\begin{eqnarray}
  && A_f(K,n)_{\text{prime}} = \frac{1}{2K} ((1+n)^K - (1+n)) \\
  && W =   (\frac{n}{2})^2(K-1)
\end{eqnarray}

$Q^-$ goes from bosonic to fermionic and from even to odd.
\begin{eqnarray}
  A_{fos^+}(K,n)-A_{bes^+}(K,n) = 
  A_{bos^+}(K,n)-A_{fes^+}(K,n) =
  -A_f(K,-n) - \frac{W}{2} \\
  A_{fos^-}(K,n)-A_{bes^-}(K,n) = 
  A_{bos^-}(K,n)-A_{fes^-}(K,n) =
  -A_f(K,-n) + \frac{W}{2}
\end{eqnarray}
The minimum total number of massless states must therefore be
\begin{eqnarray}
  -4 A_f(K,-n) = -\frac{2}{K} ((1-n)^K - (1-n)) =\frac{2}{K} (3^K - 3)
\end{eqnarray}
For $K=5$, this comes to $96$ states which is way more than the $8$
purely bosonic states with $4$ or $5$ partons that we have found in
this section.


\section{Correlation functions}

\label{sec:cor}


One of the physical quantities we can calculate nonperturbatively is the
two-point function of the stress-energy tensor. Previous calculations of
this 
correlator in this and other theories can be found
in~\cite{Antonuccio:1999iz,Hiller:2000nf,Hiller:2001qb}.
Ref.~\cite{Antonuccio:1999iz} gives results
for the theory considered here but only for resolutions
$K$ up to 6. We can now reach $K=12$.

We will show that there is a distinct behavior for even and odd $K$ in the
correlation function, just as in the energy spectrum. Then we will argue, by
taking a closer look at the data, that we have two different classes
of representations 
at finite $K$, which become identical as $K \to \infty$.


\subsection{Correlation functions in supergravity}


Let us first recall that there is a duality that relates the results for the
two-point function in ${\cal N}$=(8,8) SYM theory to the results in string
theory~\cite{Hiller:2000nf}.  The correlation function on the
string-theory side, which can be calculated with use of the supergravity
approximation, was presented in~\cite{Antonuccio:1999iz}, and we will only
quote the result here. The computation is essentially a generalization
of that given in~\cite{Gubser:1998bc,Witten:1998qj}. The main conclusion on
the supergravity side was reported in~\cite{Hashimoto:1999xu}. Up to a
numerical coefficient of order one, which we have suppressed, it was found
that 
\begin{equation}
   \bra {\cal O}(x){\cal O}(0)\ket=\frac {N_c^{\frac 32}}{g_{YM}x^5}.
   \label{two}
\end{equation}
This result passes the following important consistency test. The SYM theory
in two dimensions with 16 supercharges has conformal fixed points in both
the
UV and the IR regions, with central charges of order $N_c^2$ and $N_c$,
respectively. Therefore, we expect the two-point function of the
stress-energy tensor to scale like $N_c^2/x^4$ and $N_c/x^4$ in the deep
UV and IR regions, respectively.  According to the analysis
of~\cite{Itzhaki:1998dd}, we expect to deviate from
these conformal behaviors and cross over to a regime where the supergravity
calculation can be trusted. The crossover occurs at $x=1/g_{YM}\sqrt{N_c}$
and $x=\sqrt{N_c}/g_{YM}$. At these points, the $N_c$ scaling of \eqref{two}
and the conformal result match in the sense of
the correspondence principle~\cite{Horowitz:1996nw}.

We should note here that this property for the correlation functions is
expected {\em only} for ${\cal N}$=(8,8) SYM theory, not for the theory in
consideration in this paper. However, it would be natural to expect some
similarity between ${\cal N}$=(8,8) and ${\cal N}$=(2,2) theories.
Indeed, we will find numerically that \eqref{two} is {\em almost} true in
${\cal N}$=(2,2) SYM theory.


\subsection{Correlation functions in SUSY with 4 supercharges}


We wish to compute a general expression of the form
$F(x^-,x^+)=\bra {\cal O}(x^-,x^+){\cal O}(0,0)\ket$ where
${\cal O}$ is $T^{++}$. In DLCQ, where we fix the total momentum in the
$x^-$ direction, it is more natural to compute the Fourier transform and
express the transform in a spectral decomposed
form~\cite{Antonuccio:1999iz,Hiller:2000nf}
%
\begin{eqnarray}
  \nonumber
  \tilde F(P_-,x^+)=\frac 1{2L}\bra{T^{++}}(P_-,x^+){T^{++}}(-P_-,0)\ket 
\\
  =\sum_i \frac 1{2L}\bra 0|{ T^{++}}(P_-,0)|i\ket\e^{-iP^i_+x^+}
  \bra i|{ T^{++}}(-P_-,0)|0\ket.
\end{eqnarray}
%
The position-space form of the correlation function is recovered by
Fourier transforming with respect to $P_-=P^+ =K\pi/L$. We can continue to
Euclidean space by taking $r=\sqrt{2x^+x^-}$ to be real. The result for
the correlator of the stress-energy tensor was presented
in~\cite{Antonuccio:1999iz}, and we only quote the result here:
\begin{equation}  
  F(x^-,x^+)\equiv\bra T^{++}({\bf x})T^{++}(0)\ket =
   \sum_i \Big|\frac L{\pi}\bra 0|T^{++}(K)|i\ket\Big|^2\left(
   \frac {x^+}{x^-}\right)^2 \frac{M_i^4}{8\pi^2K^3}K_4(M_i \sqrt{2x^+x^-}),
   \label{cor}
\end{equation}
where ${\bf x}$ has light cone coordinates $x^-,x^+$,
$M_i$ is a mass eigenvalue and $K_4(x)$ is the modified Bessel
function of order 4. In~\cite{Antonuccio:1998mq} we found that the momentum
operator $T^{++}({\bf x})$ is given by
\be 
T^{++}({\bf x}) =\tr \left[(\d_- X^I)^2+\frac 12(iu^{\a}\d_-u^{\a}-i
   (\d_-u^{\a})u^{\a})\right], \quad I,\a=1,2,
\ee 
where $X$ and $u$ are the physical adjoint scalars and fermions,
respectively, 
following the notation of~\cite{Antonuccio:1998mq}.  When written in
terms of the discretized operators,  $a$ and $b$,
(Eqs.~(\ref{eq:discretized},\ref{eq:discretized2})), we find
\begin{eqnarray}
   &&T^{++}(K)|0\ket =\frac {\pi}{2L}\sum_{k=1}^{K-1} \nonumber \\
   && \quad  \left[-\sqrt{k(K-k)}
   a^{\dag}_{Iij}(K-k)a^{\dag }_{Iji}(k)+\left(\frac K2-k\right)
   b^{\dag }_{\a ij}   (K-k)b^{\dag }_{\a ji}(k)\right]|0\ket.
\end{eqnarray}

The matrix element $(L/\pi)\bra 0|T^{++}(K)|i\ket$ is independent of $L$
and can be substituted directly to give an explicit expression for the
two-point function. We see immediately that the correlator behaves like
$1/r^4$ at small $r$, for in that limit, it asymptotes to
\be 
\label{eq:smallr}
\left(\frac {x^-}{x^+}\right)^2F(x^-,x^+)=\frac{N_c^2(2n_b+n_f)}{4\pi^2r^4}
   \left(1-\frac 1K \right).
\ee
On the other hand, the contribution to the correlator from strictly
massless states is given by
\begin{equation}
 \left(\frac {x^-}{x^+}\right)^2F(x^-,x^+)=\sum_i\Big|\frac L{\pi}
   \bra 0|T^{++}(K)|i\ket \Big|^2_{M_i=0}\frac 6{K^3\pi^2r^4}.
\label{larger} 
\end{equation}
That is to say, we would expect the correlator to behave like $1/r^4$ at
both small and large $r$, assuming massless states have non-zero
matrix elements.


\subsection{Numerical results}

\begin{figure}[ht]
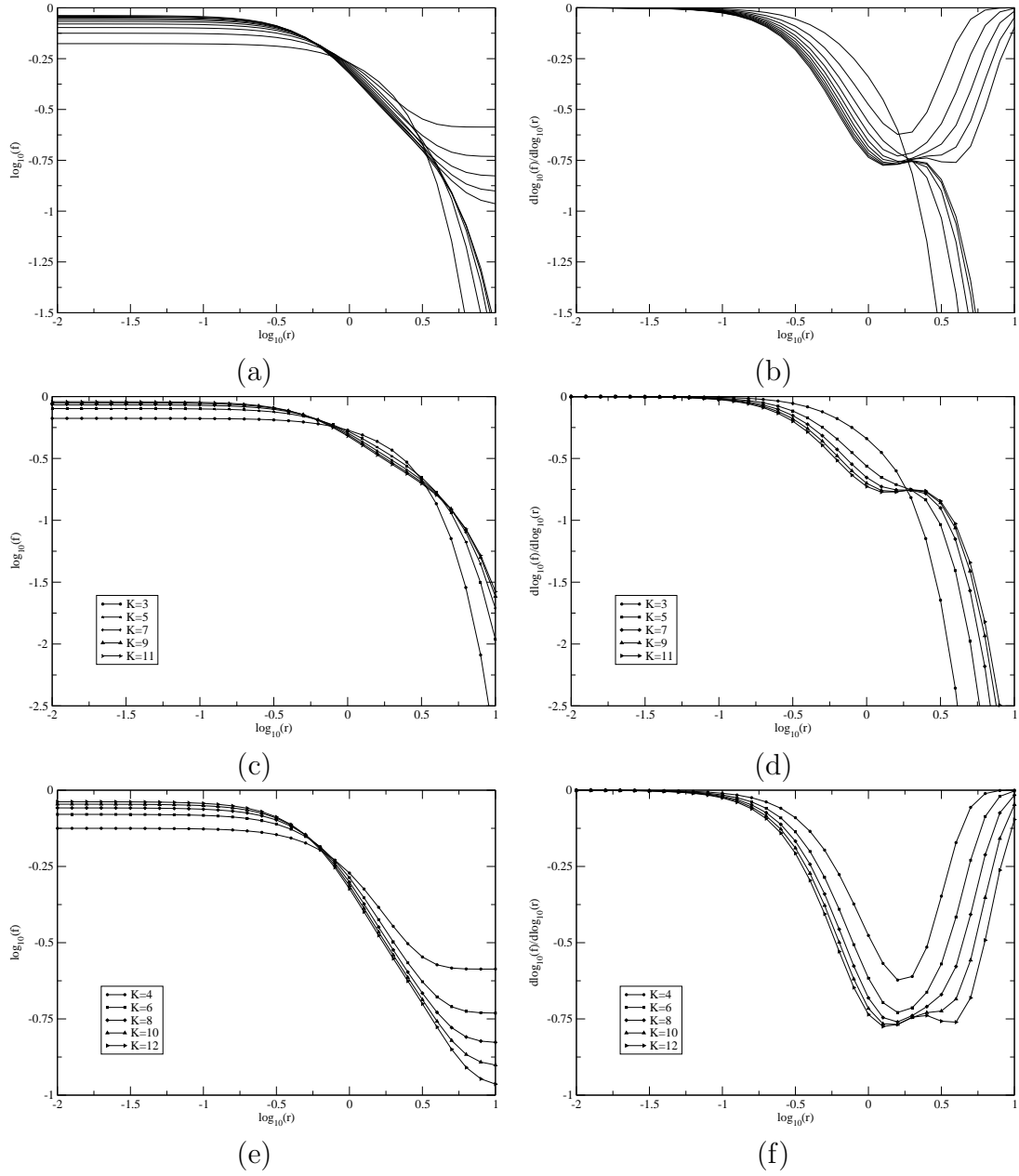

\begin{center}
\begin{tabular}{cc}
\psfig{figure=cor.eps,width=7cm} &
\psfig{figure=Dcor.eps,width=7cm}
\\
(a)&(b) \\
\psfig{figure=cor_odd.eps,width=7cm} &
\psfig{figure=Dcor_odd.eps,width=7cm}
\\
(c)&(d) \\
\psfig{figure=cor_even.eps,width=7cm} &
\psfig{figure=Dcor_even.eps,width=7cm}
\\
(e)&(f)
\end{tabular}
\end{center}
\caption{Plots of log of the scaled correlation function $f$ as a function 
of $\log_{10}(r)$ for (a) $K=3,4,\ldots 12$, (c) $K$ odd, and (e) $K$ even, 
and plots of $d\log_{10}(f)/d\log_{10}(r)$ as a function of $\log_{10}(r)$ 
for (b) $K=3,4,\ldots 12$, (d) $K$ odd, and (f) $K$ even.}
\label{cor_Dcor}
\end{figure}

To compute the correlator using Eq.~\eqref{cor}, we approximate the sum
over eigenstates by a Lanczos~\cite{Lanczos} 
iteration technique, as described in~\cite{Hiller:2000nf,Hiller:2001qb}. Only 
states with positive $R_{\a}$, $T$ and $S$ parity contribute to the correlator.
The results are shown in Fig.~\ref{cor_Dcor}, which includes a log-log plot 
of the scaled correlation function 
\be
f\equiv \bra T^{++}({\bf x})T^{++}(0)\ket 
      \left(\frac{x^-}{x^+}\right)^2\frac{4\pi^2r^4}{N_c^2(2n_b+n_f)} 
\ee 
and a plot of $d\log_{10}(f)/d\log_{10}(r)$ versus $\log_{10}(r)$, 
with $r$ measured in units of $\sqrt{\pi/g^2N_c}$. Let us discuss the 
behavior of the correlator at small, 
large, and intermediate $r$, separately in the following.

First, at small $r$, the graphs  of $f$ for different $K$ approach 0 
as $K$ increases.  This follows Eq.~(\ref{eq:smallr}) which
gives the form $f = \log(1-\frac{1}{K})$.
Second, at large $r$, obviously, the behavior is different for odd $K$, in 
Fig.~\ref{cor_Dcor}(c) and (d), and even $K$, in (e) and (f). However, the 
difference gets smaller as $K$ gets bigger, as seen in Fig.~\ref{cor_Dcor}(a). 
The reason for this is as follows. Looking at the detailed information of the 
computation of the correlator, we found that for even $K$ there is exactly 
one massless state that contributes to the correlator, while there is no 
massless state nor even an anomalously light state that makes any
contribution for odd $K$. Instead, it is the lowest massive state that 
contributes the most for odd $K$. This observation serves as another piece of 
evidence for the claim that we have two distinct classes of
representations for odd 
and even $K$.  

In the intermediate-$r$ region, for the ${\cal N}$=(8,8) theory we 
expected from Eq.~\eqref{two} that the behavior is $1/r^5$,
and in~\cite{Hiller:2000nf} we found that the correlator may be 
approaching this behavior. We indicated in~\cite{Hiller:2000nf} that 
conclusive evidence would be a flat region in the derivative of the 
scaled correlator at a value of $-1$. Our resolution was not high
enough to see this in the ${\cal N}$=(8,8) case.
Here we find such a flat region, 
indicating that the correlator in fact behaves like $1/r^{-4.75}$ for 
${\cal N}$=(2,2) SYM theory. Also, note that the region of flattening 
around $-0.75$ extends farther out as $K$ gets bigger, for both odd
and even $K$, 
implying again 
that the representations appear to 
agree as $K$ goes to infinity.
For any fixed value of $r$ the correlators for odd and even 
$K$ approach each other as $K$ increases and the flat region extends further.
This indicates
that it is only in the region of $r$ where the correlators for even and 
odd $K$ agree that we have sufficient convergence for the results to be 
meaningful. 

%

\section{Discussion} \label{sec:discussion}


To respond to the increasing interest in calculating
supersymmetric theories on a 
lattice~\cite{Cohen:2003xe,Cohen:2003qw,Sugino:2003yb,Sugino:2004qd}, we have
presented detailed numerical results for the low-energy spectrum and the 
two-point correlation function of the stress-energy tensor, using SDLCQ for 
${\cal N}$=(2,2) SYM theory in $1+1$ dimensions in the large-$N_c$ 
approximation. Our hope is that these results will serve as benchmarks 
for others to compare and check their results. 

In addition, we found an important new aspect of the SDLCQ approximation in 
this calculation.  There seem to be two distinct classes of representations for
${\cal N}$=(2,2) SYM theory, one where $S$ and $K$ have the same parity 
and one where $S$ and $K$ have opposite parity; these representations become 
identical as $K\to \infty$. We found evidence for this feature of 
${\cal N}$=(2,2) SYM theory in both the mass spectrum and the correlator. We 
also found that there are some anomalously light states that appear only in the 
sectors where $S$ and $K$ have opposite parity. We argued that the anomalously 
light states should be exactly massless, but have acquired
a tiny mass because of some impediment to having them exactly massless in the 
SDLCQ approximation. In the calculation of the correlator where only positive S parity 
contribute we found that
there is exactly one massless state that contributes to the correlator when $K$ has positive parity
and that no massless state or anomalously light  state contributes when $K$ has negative
parity. The lightest massive  state in the sector where  $K$ has negative 
parity does
contribute to  the correlator, but because the mass gap appears to close at infinite resolution
this state appears to become massless, as expected~\cite{Witten:1995im}. 

The two-point correlator of the stress-energy tensor was found to show 
$1/r^4$-behavior in the UV (small $r$) and IR (large $r$, $K$ even)
 regions as expected. 
The large
$r$ behavior for $K$ odd, on the other hand, has an exponential decay.
Surprisingly, the correlator behaves like $1/r^{4.75}$ at intermediate
values of $r$.
In ${\cal N}$=(8,8) 
SYM theory in $1+1$ dimensions, the correlator is expected to behave like
$1/r^5$ in the intermediate region, and it is interesting that ${\cal N}$=(2,2) behaves 
similarly but with a different exponent.
We were able to confirm this power law behavior with a flat 
region in the derivative of the scaled correlator.
Previously, in our calculation of the ${\cal N}$=(8,8) correlator at lower
resolutions, we were not able to find this flat region. We are hopeful that 
in the near future we may be able to conclusively confirm the $1/r^5$ behavior 
in the ${\cal N}$=(8,8) theory.  Interestingly, we also note that earlier 
results seem to indicate the same type of odd/even behavior for the
${\cal N}$=(8,8) theory.

Analytically, we investigated the properties of pure bosonic massless states 
and found the necessary and sufficient conditions  to determine their wave 
function. Then we explored some special cases to find that there are $K+1$ 
massless states of type
\[\tr[a_{A_1}^{\dag}(1)a_{A_2}^{\dag}(1)\ldots a_{A_K}^{\dag}(1)]|0\rangle,\]
where $A_i$ is a flavor index and the number in the parentheses tells how 
many units of momentum each parton carries, and that there are two massless 
states of the type
\[\tr[a_{A_1}^{\dag}(1)a_{A_2}^{\dag}(1)\ldots a_{A_{K-1}}^{\dag}(2)]|0\rangle.\] 
We also gave the formulae to count a minimum total number of massless states 
for a SYM theory which is demensionally reduced to one spatial and one time 
dimensions.

What prevents us from reaching even higher $K$ is obviously the fact that, as 
one can show~\cite{ToAppear}, the total number of basis states grows like 
$\sim (1+n)^K$, where $n$ is the total number of particle types and $n=4$ for 
${\cal N}$=(2,2) SYM theory. Our numerical results were obtained using one 
single PC with memory of 4 GB.  The problem that we now face is
that we do not have enough memory to store all the states in one PC. However, 
as we make use of a cluster of PCs and find ways to split and share the 
information among them, we are able to reach even higher $K$. This is the 
direction of our future work, with the ultimate goal being to achieve
sufficient numerical precision to detect the correspondence between $\N 8$ SYM 
theory and supergravity conjectured by Maldacena~\cite{Maldacena:1997re}. 

\section*{Acknowledgments}
This work was supported in part by the U.S. Department of Energy
and the Minnesota Supercomputing Institute.

\end{document}